\documentclass[useAMS,usenatbib]{mn2e}
\usepackage{graphicx,natbib}
\bibpunct{(}{)}{;}{a}{}{,}

\begin{document}

\sloppypar

\title[On propagation of a matter in accretion flows]{Observational evidence for matter propagation in accretion flows}

\author[M. Revnivtsev et al.]{Revnivtsev M. $^{1,2}$ \thanks{E-mail: revnivtsev@iki.rssi.ru}, 
S. Potter $^{3}$, Kniazev A. $^{3,4}$, Burenin R.$^{1}$,
\newauthor
 Buckley D.A.H. $^{3,4}$, E. Churazov $^{5,1}$\\
$^{1}$ Space Research Institute, Russian Academy of Sciences, Profsoyuznaya 84/32, 117997 Moscow, Russia \\
$^{2}$ Excellence Cluster Universe, Technische Universit\"at M\"unchen, Boltzmannstr.2, 85748 Garching, Germany\\
$^{3}$ South African Astronomical Observatory, PO Box 9, 7935 Observatory, Cape Town, South Africa\\
$^{4}$ Southern African Large Telescope Foundation, PO Box 9, 7935 Observatory Cape Town, South Africa\\
$^{5}$ Max-Planck-Institut fuer Astrophysics, Karl-Schwarzschild-str.1 , 85741, Garching, Germany
}

\date{Accepted ??, Recieved ??}

\pagerange{\pageref{firstpage}--\pageref{lastpage}} \pubyear{2010}

\maketitle

\label{firstpage}

\begin{abstract}
We study simultaneous X-ray and optical observations of
  three intermediate polars EX Hya, V1223 Sgr and TV Col with the aim
  to understand the propagation of matter in their accretion flows. We
  show that in all cases the power spectra of flux variability of
  binary systems in X-rays and in optical band are similar to each
  other and the majority of X-ray and optical fluxes are correlated
  with time lag $<1$ sec.  These findings support the idea that
  optical emission of accretion disks, in these binary systems,
  largely originates as reprocessing of X-ray luminosity of their
  white dwarfs. In the best obtained dataset of EX Hya we see that the
  optical lightcurve unambiguously contains some component, which
  leads the X-ray emission by $\sim7$ sec. We interpret this in the
  framework of the model of propagating fluctuations and thus deduce
  the time of travel of the matter from the innermost part of the
  truncated accretion disk to the white dwarf surface. This value
  agrees very well with the time expected for matter threaded onto the
  magnetosphere of the white dwarf to fall to its surface. The
  datasets of V1223 Sgr and TV Col in general confirm these findings,
  but have poorer quality.
\end{abstract}

%

\section{Introduction}

Accretion is the main source of energy for a wide variety of
astrophysical objects, from pre-main sequence stars through white
dwarf, neutron star and black hole binaries up to supermassive black
holes in centers of galaxies. The matter in accretion disks in these
systems gradually moves towards the compact object, extracts
gravitational energy and produce broad band emission spectra.  In
spite of general understanding of formation of emission in accretion
disks \cite[see e.g][]{ss73,done07} and temperature distributions
over the disk \cite[see e.g.][]{horne85}, the very fact that the
matter travels outside-in is not an easy thing to verify even though
it is essential for the whole accretion process.

Virtually the only way to verify this movement of the matter in the
accretion flow is to study the time variability of its emission at
different parts of the flow (except for, may be, the radial velocity 
component of the emission lines). Indeed, matter travels from the outer
parts of the accretion flow towards the central compact object and
thus any time variations of the mass tranfer rate in the flow should
be transported inwards (though with possible smearing due to influence
of viscosity).

This idea together with the assumption that all accretion flow radii
generate their own additional noise at characteristic frequencies,
corresponding to their dynamical times is an essense of the {\sl model
  of propagating fluctuations} \citep[see
  e.g.][]{lyubarskii97,churazov01,uttley01,kotov01,arevalo06,revnivtsev09,revnivtsev10}. The
shortest time scales in this model are introduced at the smallest
radii, the largest time scales are introduced at outer parts of the
accretion disks. The inner parts of the disk add their noise to the
mass accretion rate coming from the outer parts in a multiplicative
way, thus naturally producing the observed linear relation between the
amplitude of the fluctuations and the time averaged flux \cite[see
  e.g.][]{lyutyi87,uttley01} and log-normal distribution of
instantaneous values of fluxes \cite[see
  e.g.][]{uttley05,revnivtsev08}. The model implies that there should
be a definite time lag between variabilities of emission of outer and
inner parts of the accretion flow. Note than here we can compare only
variabilities at long time scales, because short time scale
fluctuations are absent at the outer parts of the disk.

This prediction is not easy to check because of different complications.
 For example, in the case of 
galactic neutron star and black holes binaries it is relatively easy to observe the 
variability of the mass accretion rate in the innermost parts of the flow -- in the 
X-ray energy band. At the same time, 
the mass accretion rate variations at the outer parts of the accretion disk are almost invisible 
to us because the internal energy release of accreting matter at these radii is negligible 
in comparison with the energy absorbed by the disk from the illuminating X-ray flux 
\cite[e.g.][]{dubus99}. Therefore, the optical emission of these systems is largely 
determined by the reprocessing of their X-ray luminosity \cite[e.g.][]{vanparadijs94} 
and thus does not provide us information about internal mass accretion rate variations 
at these radii. In accretion disks around supermassive black holes the characteristic 
time scales can be as large as years and tens of years, therefore requiring large 
monitoring campaigns, rarely avaiable \cite[see, however,][]{edelson96,desroches06,doroshenko09,arevalo09}.

One of the best available examples today of propagation of mass accretion rate 
fluctuations (flickering) in the accreting systems can be found among dwarf novae -- 
accreting non magnetized white dwarfs. In the work  of \cite{pandel03} 
it was shown that the X-ray emission of accreting white dwarf VW Hyi in quiescence 
is delayed with respect to its UV emission with $\Delta t\sim$100 sec. 
It is assumed that in this system the optically thick accretion disk, emitting 
UV radiation, ends at some distance from the white dwarf while the X-ray emission 
originates at the WD surface. The observed time lag was interpreted by authors as 
a time for the matter to travel from the inner parts of the optically thick accretion disk to the WD surface. 

Due to rather uncertain issue about the disk truncation in case of dwarf novae in quiescence 
it is reasonable to look for better observational evidences of matter propagation in accretion flows.
 For this purpose we have selected luminous intermediate polars (IPs) -- magnetized white dwarfs, in which 
the accretion disk is truncated very close to the WD surface (although some IPs maybe discless systems), 
but which, nevertheless, certainly have geometrically distinct regions generating the outgoing radiation. 
In intermediate polars X-rays originate close to the surface of WD, while the optical emission is mainly 
generated by the optically thick accretion disk or accretion curtains \cite[see e.g.][]{hellier87,patterson94,hellier95}. The optical emission of 
the disk can be powered either by its internal dissipation or, by a reprocessing of the X-ray emission, 
coming from the central object \cite[see e.g.][]{beuermann04}

In cases of intermediate polars we have several advantges: 1) we know that the accretion disk 
is certainly truncated at some distance from WD because we see X-ray pulsations, 2) we can make 
an estimate of the innermost radius of this accretion disk judging from the shape of the power 
spectra of their time variability \cite[see e.g.][]{revnivtsev09,revnivtsev10}.
High mass accretion rates in these systems ensure that the white dwarf magnetosphere is not 
large and thus the internal energy release in the accretion disk is not completely negligible 
in comparison with the illuminating X-ray flux from the WD.

We have performed a set of simultaneous observations of EX Hya, V1223
Sgr and TV Col in X-ray and in optical spectral band with RXTE/PCA and
the 1.9m telescope of the South African Astronomical Observatory. In
this paper we present results of this campaign.

\section{Observations and data reduction}

\subsection{RXTE data}

Our sample includes three of the brightest intermediate polars of the
southern hemisphere (in order to ensure simultaneous observations with
South African Astronomical Observatory): EX Hya, V1223 Sgr and TV
Col. These sources were observed by the RXTE \citep{rxte} during
approximately 20 ksec each in April 2010. More detailed log of
simultaneous RXTE-SAAO observations is presented in Table 1.

\begin{table*}
\caption{Log of X-ray and optical observations, used in the paper}
\tabcolsep=4mm
\begin{tabular}{c|ccc|ccc}
       &\multicolumn{3}{c}{RXTE}&\multicolumn{3}{c}{SAAO}\\
Source &   Obs.ID &Start time&Exp., ksec& Start time (MJD)&Exp., ksec&Filters\\
\hline
EX Hya &   95305-02-01-00&55301.828& 3.3&   55301.818  &5.1&R,U\\
EX Hya &   95305-02-04-00&55303.986& 3.2&   55303.973  &4.7&R,U\\
EX Hya &   95305-02-02-02&55304.836& 3.2&   55304.822  &4.6&R,U\\
V1223 Sgr& 95305-01-02-05(10)&55305.066&2.7&   55305.072  &8.2&R,U\\
TV Col&  95305-03-01-00 & 55301.752   &2.0&   55301.750  &4.7&I,B\\
TV Col&  95305-03-02-00 & 55303.764   &3.4&   55303.751  &3.6&R,U\\
TV Col&  95305-03-02-01 & 55304.744   &4.6&   55304.739  &4.9&R,U\\
\hline
\end{tabular}
\end{table*}

Data were analized with tasks of HEASOFT package, version V6.5. RXTE/PCA background was estimated with the help of model, appropriate for faint sources "CMFAINT\_L7". Lightcurves of sources were extracted from data of the first layer of PCU2 in energy band 3-15 keV, maximizing the signal to noise ratio. All lightcurves were background subtracted for the analysis.

\subsection{SAAO data}

For aquiring optical lightcurves we used the recently commissioned HIgh speed Photo-POlarimeter (HIPPO; \citealt{potter10}) on the 1.9m telescope of the
South African Astronomical Observatory during the nights beginning
15th, 17th and 18th April 2010 (details of the observations is presented in Table 1). HIPPO is a 2 channel instrument
capable of simultaneous two filtered photo-polarimetry. None of the
targets showed statistically significant polarization in any filter
and consequently the photometry is reported here only. Data
reduction proceeded as outlined in \cite{potter10} and binned to 1
second time resolution. Absolute timing was maintained via the
observatory's time service which is phased by a GPS receiver.

\section{Power spectra}

Observations clearly show aperiodic variability of fluxes of all sources both in X-ray and in optical spectral bands. Among the obtained datasets, observations of EX Hya have the best quality (the source is brighter than TV Col in X-rays, and the length of overlaping observations is much larger than for
V1223 Sgr), therefore we will concentrate below on the case of EX Hya, while presenting the similar results (if statistics allows us) for other sources.

\begin{figure}
\includegraphics[width=\columnwidth,bb=22 177 568 700,clip]{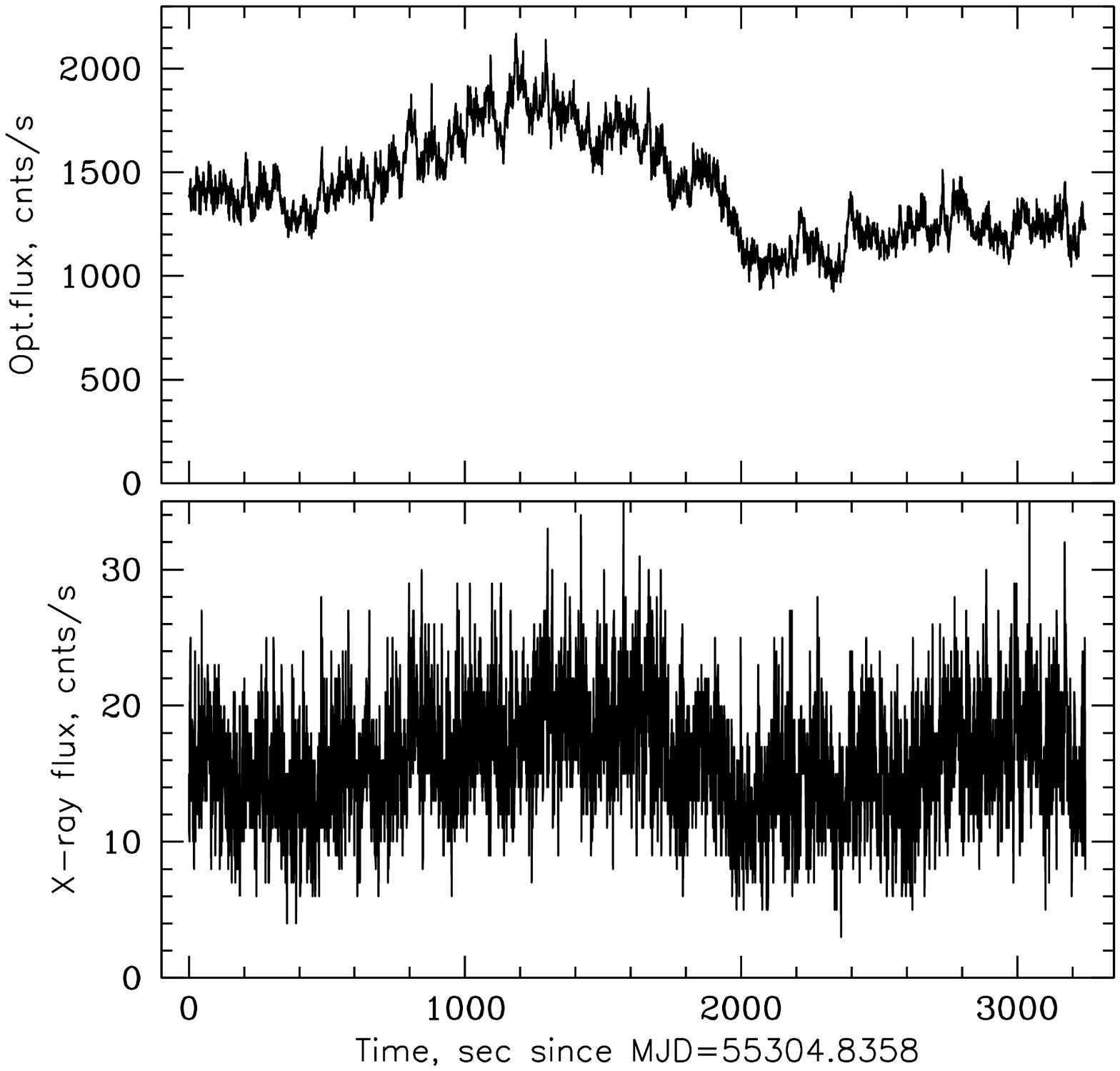}
\caption{Part of the lightcurve of EX Hya, recorded on April 18, 2010.}
\label{lcurve}
\end{figure}

An example of lightcurves of EX Hya, observed simutaneousely with RXTE/PCA and SAAO is shown in Fig.\ref{lcurve}. Close similarity between curves is clearly seen. 

\begin{figure}
\includegraphics[width=\columnwidth,bb=22 177 568 700,clip]{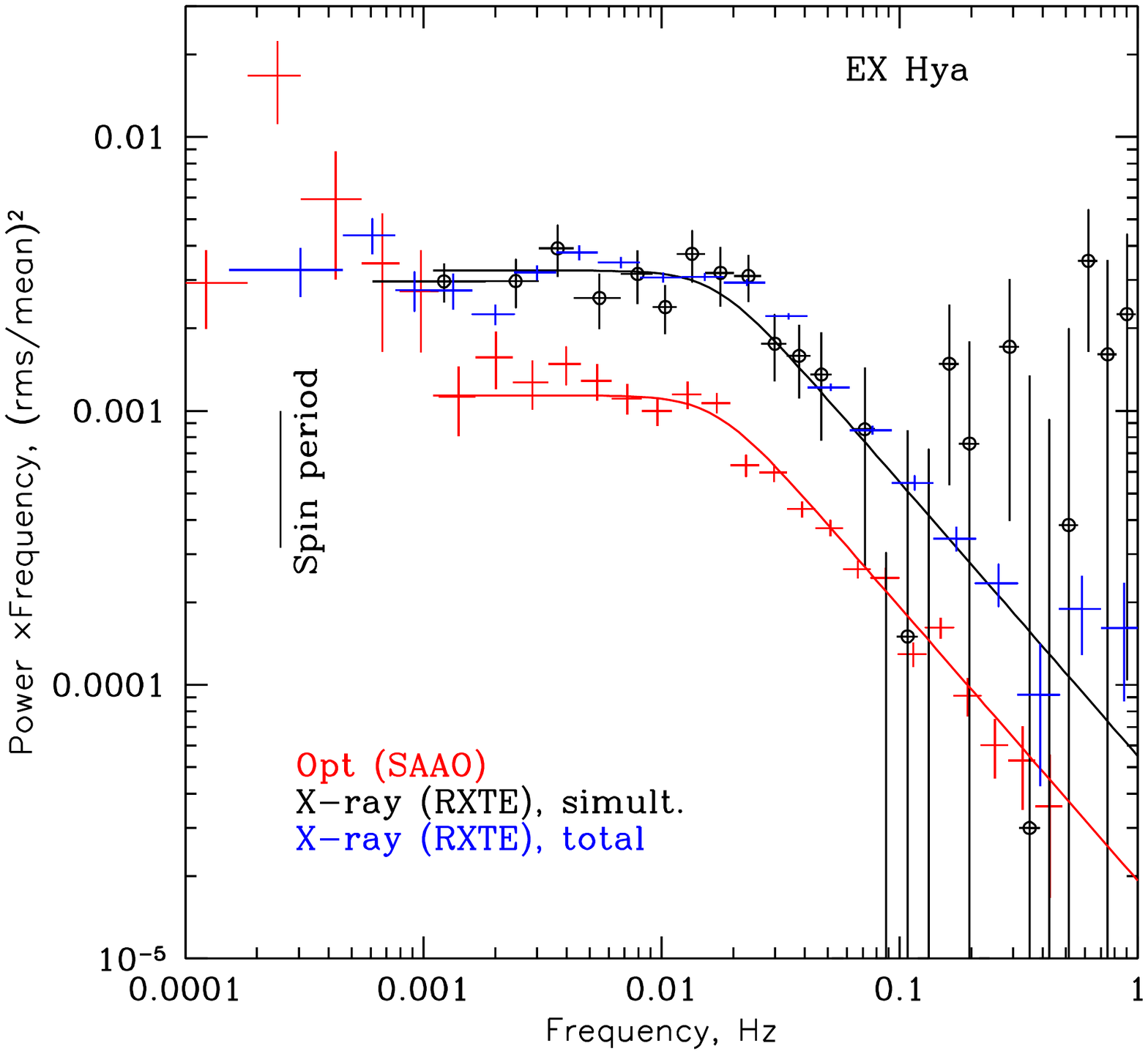}
\caption{Power spectra of variability of EX Hya in X-ray and in optical spectral bands. Two power spectra of X-ray variability represent data of April 2010 (black circle) and all data of RXTE archive (blue crosses). Solid curves show the simplest analytical approximation of power spectra (see text). The frequency of the break $f_0=1.7\times10^{-2}$ Hz.}
\label{power_ex_hya}
\end{figure}

The power spectra of variability of the source at these energies are very similar to each other (see  Fig.\ref{power_ex_hya}). 

The shape of the power spectra can be adequately described by a simple analytical model (following \citealt{revnivtsev10})

$$
P(f)\propto f^{-1}\left[1+\left({f\over{f_0}}\right)^4\right]^{-1/4}
$$

, describing smooth break between the slope of the power spectrum at low frequencies $P(f)\propto f^{-1}$ and at high frequencies $P(f)\propto f^{-2}$. Frequencies of the break in optical in X-ray data, collected in April 2010, are $f_{0,o}=(1.5\pm0.1)\times10^{-2}$ Hz and $f_{0,x}=(1.7\pm0.3)\times10^{-2}$ Hz, correspondingly. If we fit the power spectrum of variability of X-ray flux of EX Hya, 
averaged over all observations in RXTE archive (1996-2010, exposure time $\sim$200 ksec), we 
obtain the break frequency $f_0=(2.1\pm0.1)\times10^{-2}$ Hz. 

Assuming that the break frequency corresponds to the frequency of Keplerian rotation at the 
boundary of the magnetosphere  $f_0\approx \sqrt{GM_{\rm WD}/R_{\rm in}^3}/2\pi$ \cite[see evidences 
for this statement in][]{revnivtsev09} we can estimate the inner radius of the accretion disk in 
EX Hya. We adopt the mass of the WD in EX Hya $M_{\rm WD}=0.79~M_\odot$ \citep{beuermann08}, and 
thus its radius (using \citealt{nauenberg72})  $R_{\rm WD}\sim7\times10^{8}$ cm.  The estimate of 
the innermost radius of the disk from the value of the break frequency is $R_{\rm in}\sim 1.9\times10^{9}$ cm, 
or $\sim2.7 R_{\rm WD}$. 

In fact, it is likely that the transition between the accretion disk and the WD magnetosphere is not a simple perfect circle and forms something like accretion curtains \cite[see e.g.][]{hellier87,hellier95}. Therefore it would be reasonable to say that the position of the break in the power spectrum measures the position of these transition regions.
It is remarkable to note note that estimates of distance of these accretion curtains from the WD surface made from completely different physical arguments, i.e. $R_{\rm in}\sim 1-2\times10^{9}$ cm from  analysis of emission line profiles \citep{hellier87} and $R_{\rm in}\sim 1.5\times10^{9}$ cm from analysis of spin modulated eclipses of the emission region \citep{siegel89} very close to our estimates.

This value of the radius of the transition region (or size of the WD magnetopshere) 
tells us that it is truncated much below the corotation radius. Depending on details of coupling of the WD 
magnetosphere to the accretion disk it might lead to a certain spin up of the 
WD rotation. Note, that the white dwarf in EX Hya is indeed spinning up \cite[see the 
latest measurements in][]{mauche09}.

\section{Time lags}

Similarity of optical and X-ray power spectra variability of EX Hya is naturally predicted by the model of propagating fluctuations. Variations of the mass accretion rate, flowing through the inner part of the accretion disk (which creates optical/UV emission), results in a modulation in the mass accretion rate at the WD surface, thus generating variable X-ray flux. If the optical/UV light is powered mainly by the internal dissipation in the disk, then X-ray emission variations should lag the optical variation by a matter travel time in the magnetosphere 

$$
\Delta t \sim {R_{\rm in}-R_{\rm WD}\over{\sqrt{GM_{\rm WD}/R}}}\sim 5~\textrm{sec}.
$$

\begin{figure}
\includegraphics[width=\columnwidth]{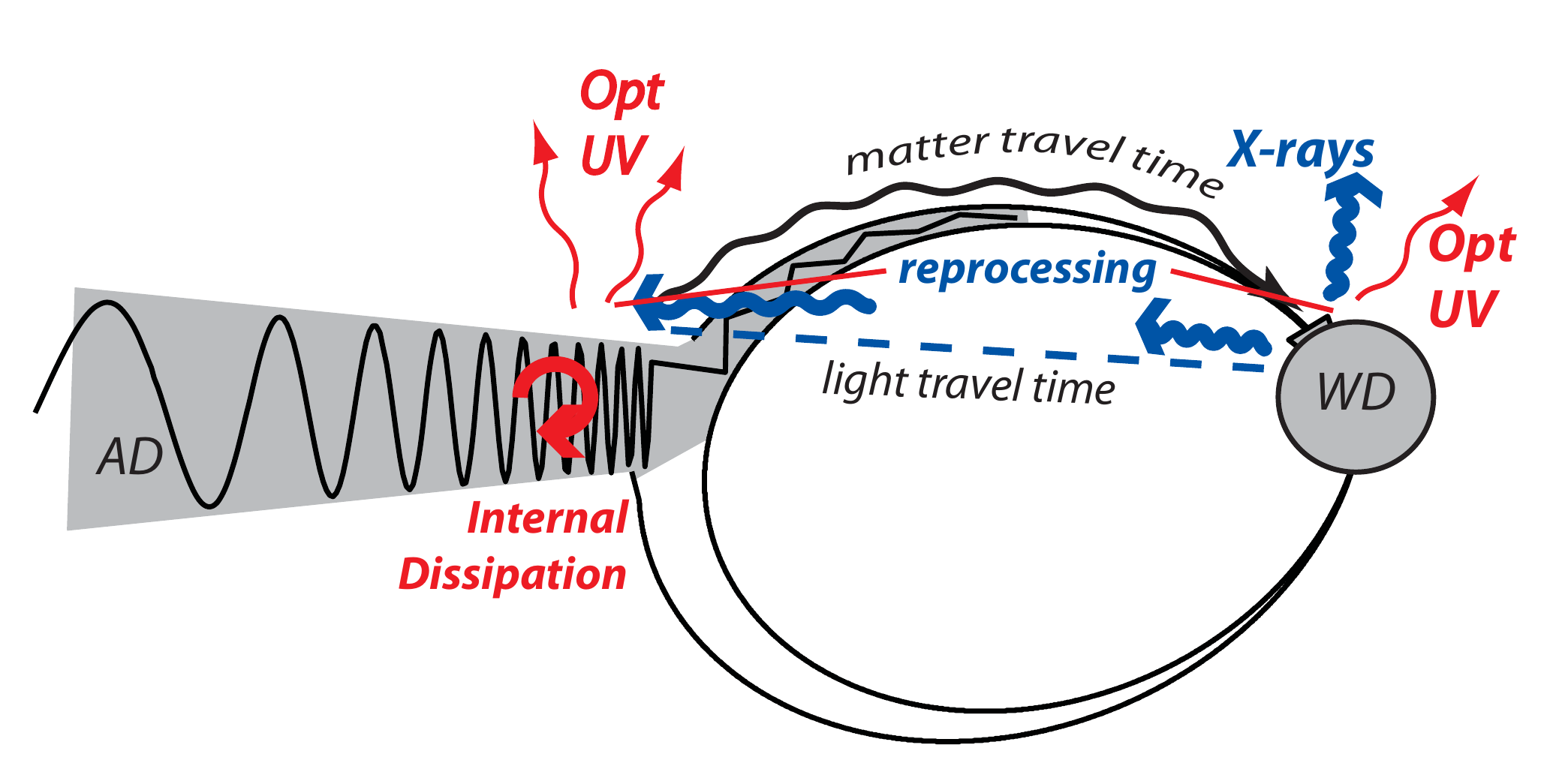}
\caption{Scheme of accretion flow in intermediate polars in the framework of the model of propagating fluctuations. The variable mass accretion rate with broad band variability (created in extended accretion disk) modulates the optical emission emerging from innermost parts of the accretion disk at the boundary of the magnetosphere, and then, after matter travel time, modulates the X-ray flux from WD surface. The X-ray flux, in turn, illuminates the inner part of the disk, which then create optical variations, in line with variations of X-ray flux.}
\label{scheme}
\end{figure}

More accurately, if we will assume, that the matter is accelerating from zero velocity at $R_{\rm in}$ towards the white dwarf and moves radially, then 
$$
\Delta t= \sqrt{ {R_{\rm in}^3\over{2GM_{\rm wd}}}} \int_{1}^{R_{\rm wd}/R_{\rm in}}{{u^{1/2} du \over{(1-u)^{1/2}}}}~,\textrm{where~} u={R\over{R_{\rm in}}}
$$

or

\begin{eqnarray*}
\Delta t= \sqrt{ {R_{\rm in}^3\over{2GM_{\rm wd}}}} & \times\\
\left[  {\pi\over{2}} - \arcsin\left(\sqrt{R_{\rm wd}\over{R_{\rm in}}}\right) + {1\over{2}}\sin\left( 2 \arcsin(\sqrt{R_{\rm wd}\over{R_{\rm in}}}) \right) \right]\\
\end{eqnarray*}

, which for our parameters is $\sim$8 sec. 
However, if the reprocessing of the X-ray emission plays a dominant role in heating of the accretion disk (note that surface of WD is heated by X-rays anyway), 
then the optical emission will lag the X-rays by a light crossing time 
$\Delta t \sim (R_{rm in}-R_{\rm WD})/c \sim 6$ msec.
Generation of this variability pattern is schematically shown in Fig.\ref{scheme}.

\begin{figure}
\includegraphics[width=\columnwidth,bb=22 177 568 700,clip]{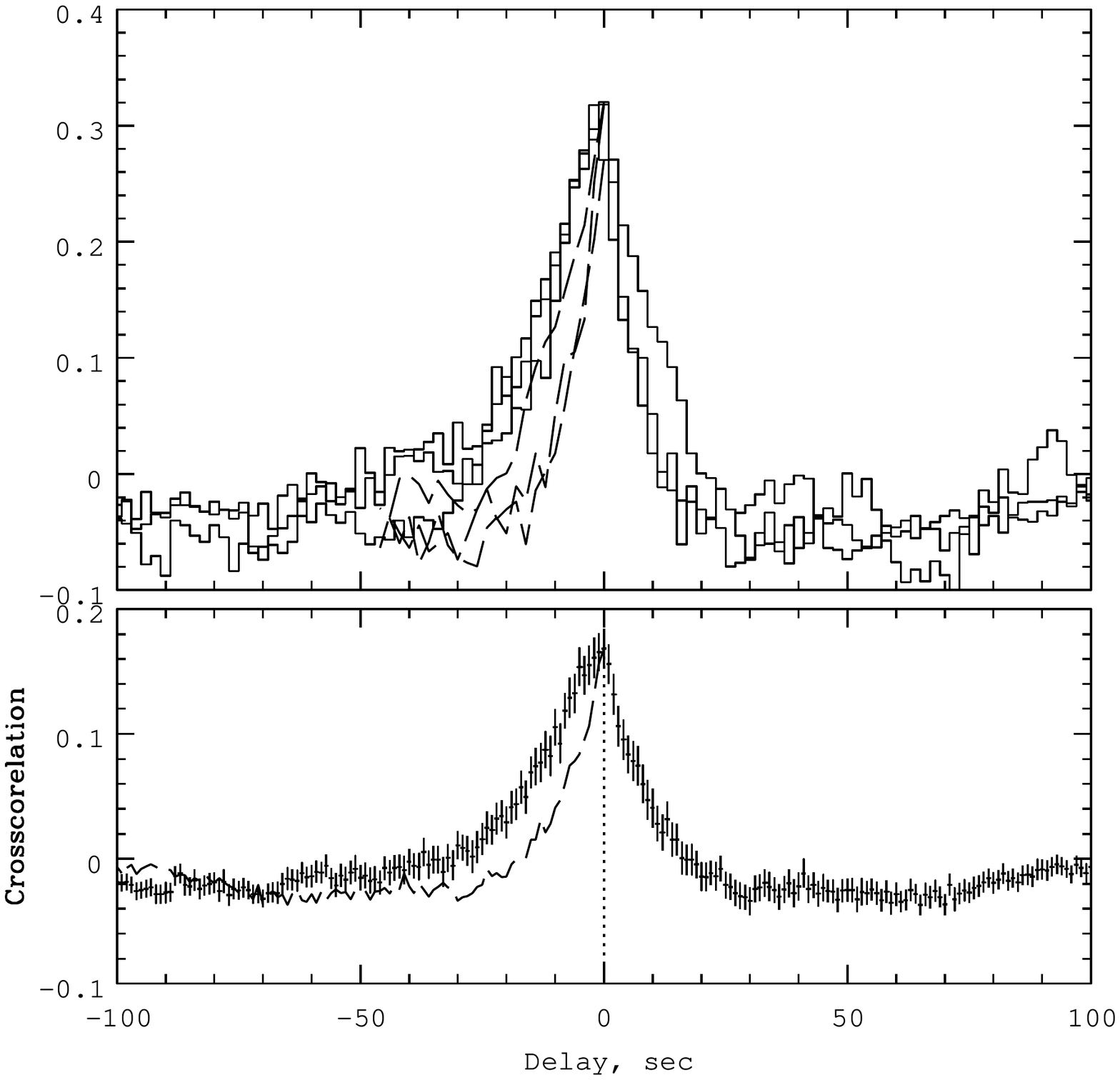}
\caption{Cross-corelation of optical (R-band) and X-ray lightcurves of EX Hya. Positive delay means that 
optical lightcurves lags the X-rays. Dashed curves on negative delays show the mirrorred cross-corelation 
at positive delays, demonstrating its skewness.  Error bars on lower panel shows the rms variations of 
the value of the cross-corelation at any given lag in different segments of the total lightcurve.
Lower panel shows cross-corelation, averaged over all aquired data, upper panel shows cross-corelations 
for three parts of the dataset separately.}
\label{cc}
\end{figure}

In both cases the flux variability of the source in these spectral bands should be closely correlated.
This is indeed observed. The curves are strongly correlated and the peak of the cross-corelation is at $\Delta t\sim 0$ (see Fig.\ref{cc}). This directly shows that the variable optical emission of EX Hya is mainly powered by reprocessing of X-rays (the light crossing time lag $\sim$6 msec can not be detected with time resolution of our datasets).

However, it is seen on Fig.\ref{cc} that the cross-corelation obviousely is not symmetric with respect to zero -- there is much more correlation at negative delays (optics leads X-rays), than on positive.
This indicates that we do see some part of the internal dissipation in the disk and its variability leads the X-rays.

\begin{figure}
\includegraphics[width=\columnwidth]{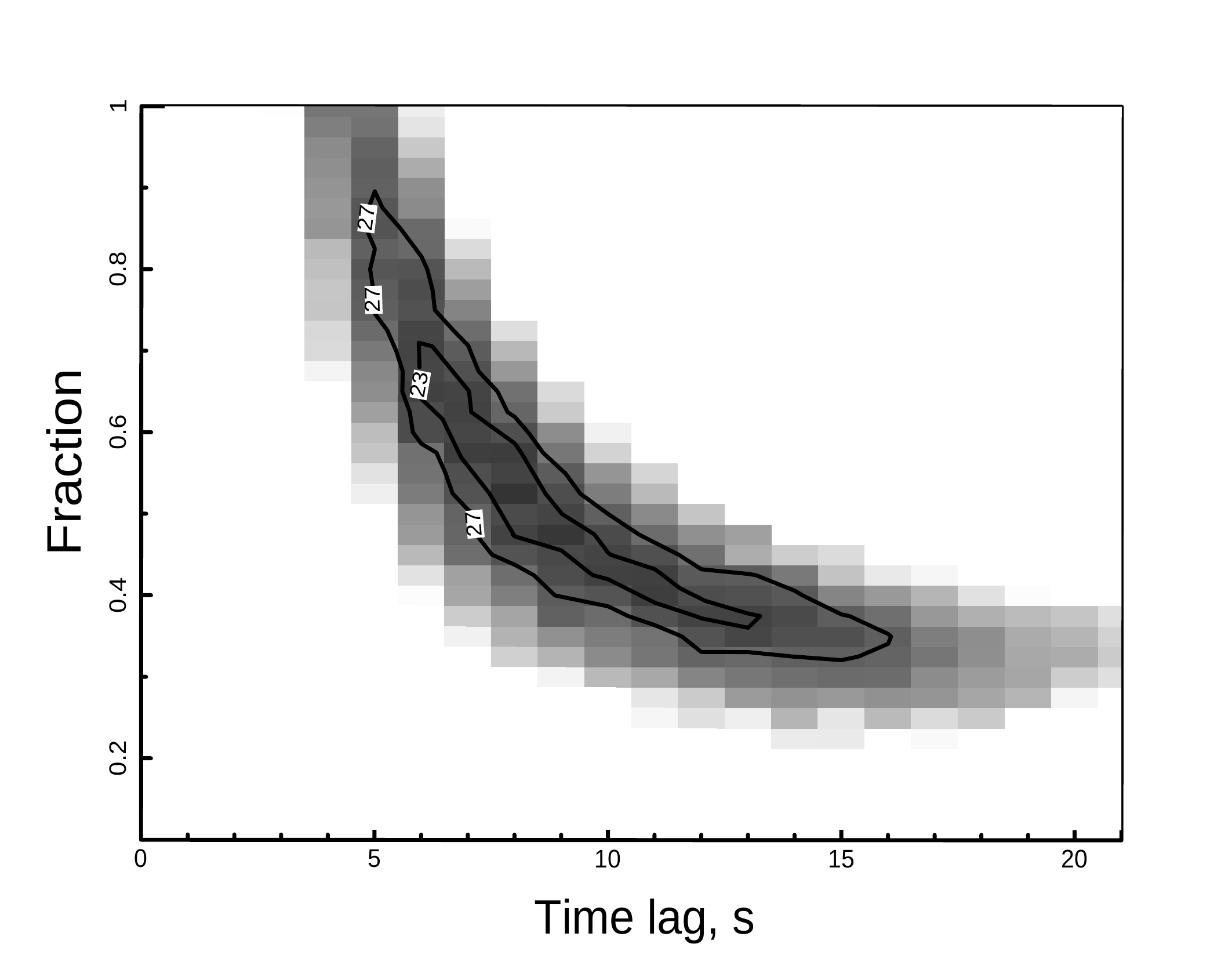}
\includegraphics[width=\columnwidth,bb=34 177 568 510,clip]{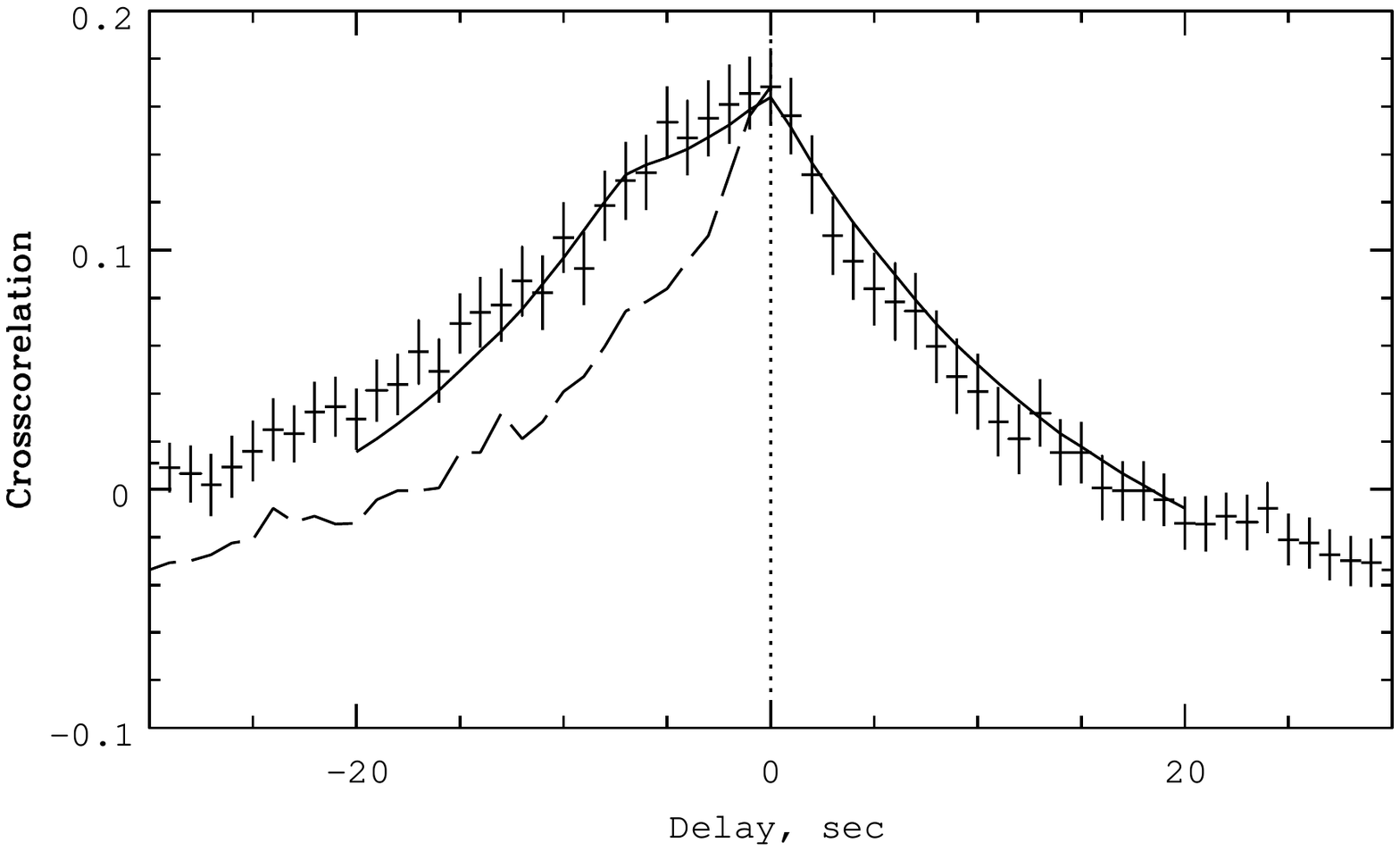}
\caption{
{\sl Upper panel}: the contour plot of the $\chi^2$ values over $\Delta t$ and $A$ values. 
$\Delta t$ is the time lag between a part of the optical curve, which is preceeding the X-ray lightcurve, 
and the value $A$ is its fractional contribution to the total optical variability. The lowest value of 
the $\chi^2$ in this plot is 20.5 for 39 degrees of freedom (however, one should keep in mind that the 
neighboring values of the cross-corelation are not statistically independent). Numbers on contours show 
the appropriate values of the formal $\chi^2$. {\sl Lower panel}: the best fit approximation of the 
observed cross-corelation function (crosses) with the cross-corelation function obtained from 
simulated lightcurves (solid curve). Here we adopted the time lag 7 seconds (part of the optical 
curve leads X-rays) and the fractional contribution of the lagged part to the total variability 
of the optical curve is 0.5. The dashed curve is the mirrorred observed cross-corelation at 
positive delays, which is shown to demonstrate the skewness of the observed cross-corelation function.
}
\label{ccfit}
\end{figure}

In order to demonstrate this point we have simulated the X-ray and
optical lightcurves, and compared their cross-corelations with the
observed one. We have simulated the X-ray lightcurve $X(t)$, which
is suppossed to be representing mass accretion rate variations on WD
surface, as a curve, whose power spectrum has a shape
measured by us (see above). Then we have simulated the optical curve
$O(t)$. This curve consists of two parts, one is having zero time lag with respect to
X-ray curve (simple reprocessing of the illuminating X-ray flux) and another, which is preceeding the X-ray curve due to 
finite matter travel time from the place of generation of the optical emission to the WD surface.
Therefore, the curve $O(t)$ was modelled as a sum of two copies of the simulated X-ray lightcurve with a range of
time lags between them ($\Delta t$) and fractional contribution of the
delayed component given by $A$: $O(t)=(1-A)\cdot X(t)+A\cdot
X(t+\Delta t)$. The cross-correlation of the resulted curves was compared
with that obtained from observations in range of delays $[-20,+20]$
seconds. The $\chi^2$ contour plot with different values of the time
lag and the fractional contribution of the delayed curve is shown on
upper panel of Fig.~\ref{ccfit}.  The minimum of the formally
calculated $\chi^2$ is approximately $20.5$ for 39 degrees of freedom
(41 data points and 2 parameters), but we should keep in mind that the
neighboring values of the cross-corelation are not statistically
independent because they use almost the same samples of observed points on lightcurves, 
therefore the face values of the $\chi^2$ can not be used
to calculate true statistical significances. If we try to rescale the
obtained minimum of the $\chi^2$ value to the number of d.o.f
(assuming, that the fit is good), the formal $1\sigma$ confidence
intervals on the parameters would be: $\Delta t=7\pm1$ sec and
$A=0.5\pm0.05$.  It is remarkable, that the lag between X-ray and
optical data perfectly agrees with the estimate of the matter travel
time in the magnetosphere of EX Hya (see above).

\section{V1223 Sgr and TV Col}

The quality of the datasets for V1223 Sgr and TV Col are somewhat worse,
therefore we cannot repeat in detail the analysis which we have
done for EX Hya. However, we do see the same similarities between power spectra in the X-ray and optical
bands, and we do see significant correlation between them.

We only would like to mention some peculiarity in the power spectrum
of variability of X-ray flux of TV Col during our observations in
2010. The power clearly has some excess (quasiperiodic oscillation,
QPO) at frequencies around $f_{\rm QPO}\sim1.6\times10^{-2}$ Hz with
an amplitude of $5\pm1$\%. On the power spectrum obtained from all the
data in the RXTE archive this excess is not so narrow, indicating that
it might be either a transient phenomenon, or a phenomenon with a
floating centroid frequency. It is interesting to note that this QPO
is located close to the frequency of the break in the power spectrum
($f_{\rm break}\sim5.1\times10^{-2}$ Hz in this case). Such behavior
is very similar to that of power spectra of accreting magnetized
neutron stars/X-ray pulars (e.g. Cen X-3 or 4U1626-67, see
\citealt{revnivtsev09}). It is likely that this QPO might be connected
with instabilities at the boundary of the compact object
magnetosphere.

\begin{figure}
\includegraphics[width=\columnwidth,bb=22 177 568 700,clip]{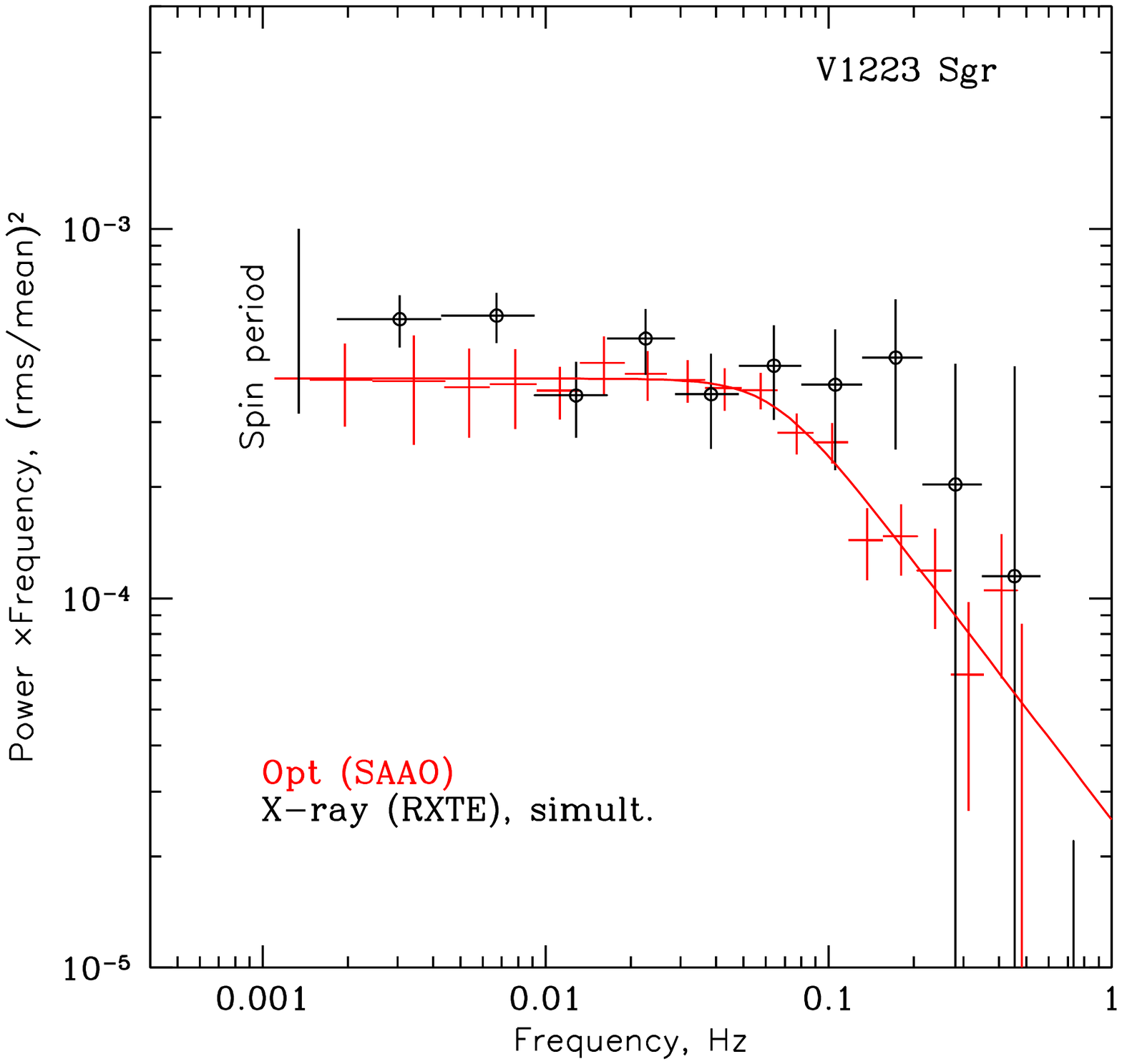}
\includegraphics[width=\columnwidth,bb=22 177 568 700,clip]{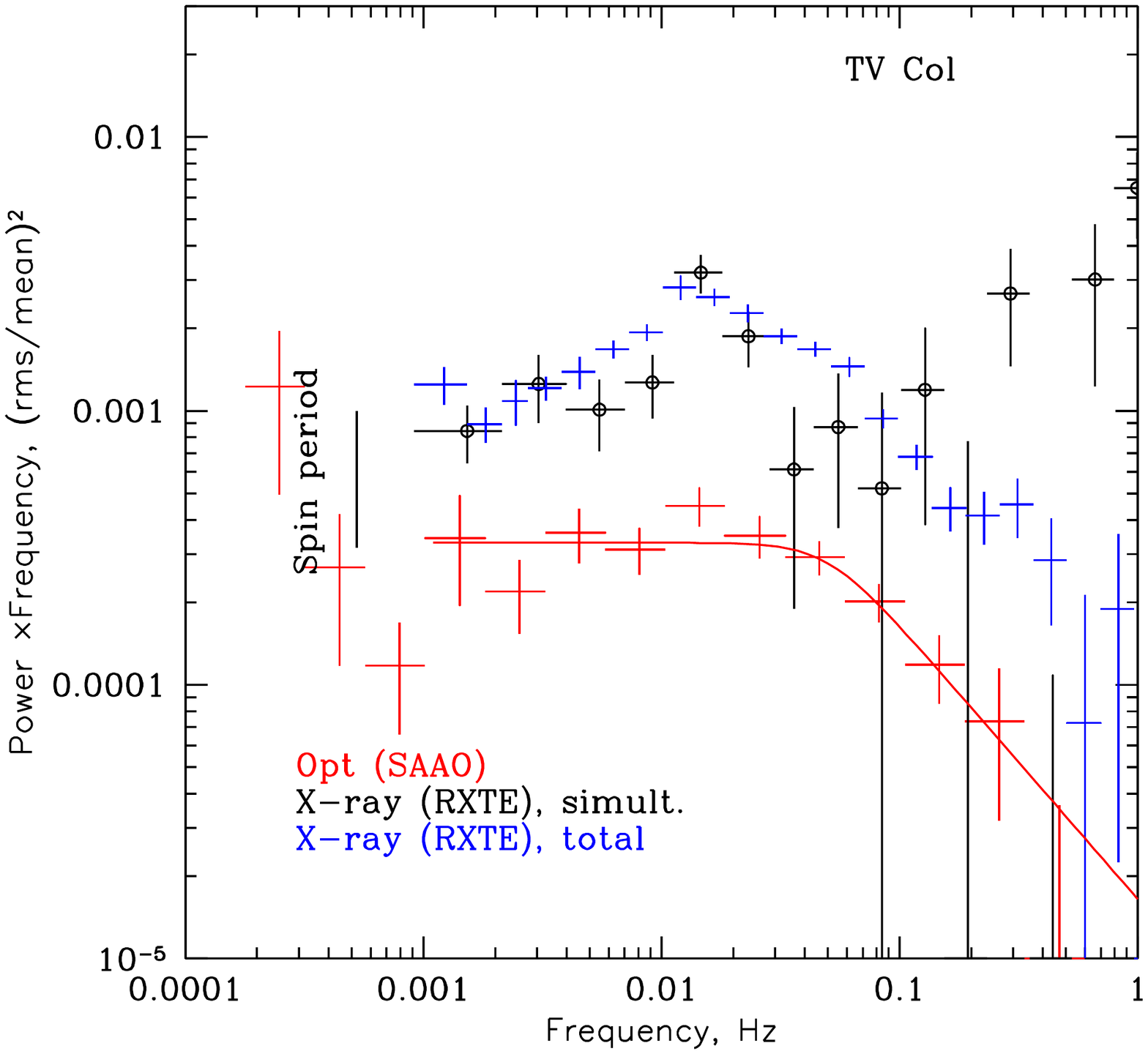}
\caption{Power spectra of variability of V1223 Sgr (upper panel) and TV Col (lower panel) in X-ray and in optical spectral bands. Open circles denote the power spectrum of X-ray flux of sources, recorded in 2010, blue crosses show the power spectrum of TV Col, averaged over all data in RXTE archive. Solid curves show the simplest analytical approximation of power spectra, measured in optical spectral band during our observational campaign.
For V1223 Sgr the break frequency $f_0=7\times10^{-2}$ Hz, for TV Col $f_0=5\times10^{-2}$ Hz.}
\label{power_v1223}
\end{figure}

\begin{figure}
\includegraphics[width=\columnwidth,bb=22 177 568 700,clip]{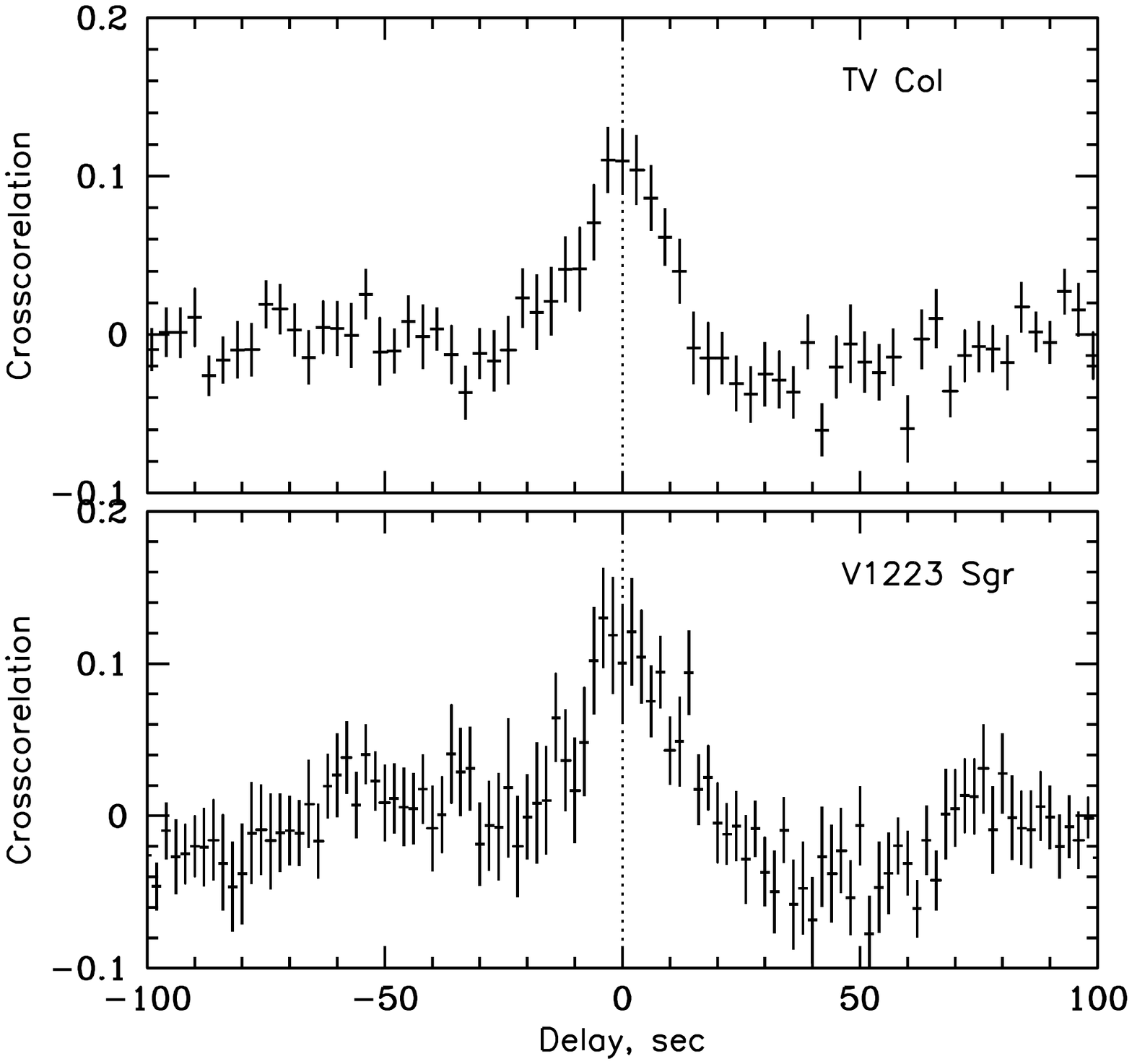}
\caption{Cross-corelation of optical and X-ray lightcurves of V1223 Sgr and TV Col, observed by SAAO and RXTE in April 2010.}
\label{cc_v1223}
\end{figure}

Correlations also peak at $\sim$zero timelag (Fig.\ref{cc_v1223}) but they are more noisy than that of EX Hya.

It should be noted, that the shape of the power spectrum of the V1223 Sgr data from 2010 (reported here) has a break
at Fourier frequency, which is significantly higher than
that, detected in the power spectrum of its variability recorded in 2008 (Revnivtsev et al. 2010) 
and that we have seen in the case of EX Hya. This indicates that
the inner accretion disk radius has decreased since the 2008 observations. Adopting this new inner radius of the
accretion disk $R_{\rm in}\sim (GM_{\rm WD})^{1/3}/(2\pi
f_0)^{2/3}\sim9\times10^{8}$ cm (here we adopted $M_{\rm
  WD}=1M_\odot$) we can estimate that the lag due to matter travel
time in the magnetosphere of V1223 Sgr should be small, $\Delta t\la
1$ sec and thus, undetectable with time resolution of our optical data
(1 sec).

\section{Summary}

We have analized the sets of simultanous observations of the X-ray brightest southern 
intermediate polars EX Hya, V1223 Sgr and TV Col. The best avaiable dataset of EX Hya allowed 
us to obtain the results summarized below. Datasets of sources V1223 Sgr and TV Col are poorer quality, 
but they demonstrate the same properties (as far as the statistics allow) of power spectra of their variability 
and cross-corelation functions.

\begin{itemize}
\item The power spectra of the flux variability of EX Hya in optical and X-ray spectral bands are very 
similar to each other and have a break in the slope $P(f)\propto f^{-1}$ to $P(f)\propto f^{-2}$ at the 
frequency $f_0\sim0.02$ Hz. Following \cite{revnivtsev09,revnivtsev10} we relate this break with the 
transition of the matter from the disk flow at larger distances from the WD to the magnetospheric flow 
closer to the WD. The break in the power spectrum is at much higher frequencies than the WD spin frequency, 
thus indicating that the disk ends within the corotation radius.
\item X-ray and optical lightcurves are strongly correlated with peak of cross-corelation function around 
zero timelag. We interpret this as a sign of reprocessing of X-ray light at surfaces of the optically 
thick accretion disk, the accretion curtains and the WD.
\item However, we detect a clear and stable asymmetry of the X-ray-optical
  crosscorrelation function, which indicates that at least some part of the optical variability leads the
 X-ray
  variability. We measure the the time lag between these variabilities
  $\Delta t\sim 7$ sec, which is consistent with the travel time of
  matter from the inner radius of the accretion disk (or accretion curtains), along the
  magnetosphere, to the white dwarf surface. We interpret this as a clear sign 
of the propagating fluctuation in the accretion flow. In
  this particular case we estimate that approximately $50$\% of the
  optical variability is preceeding the variability of the X-ray
  lightcurve, indicating a significant contribution of the internal
  energy dissipation in the disk to the total energy budget of the
  inner part of the optically thick accretion disk.
\end{itemize}

\section*{Acknowledgements}
Authors thank Coel Hellier for useful comments about accretion curtains in intermediate polars.
  This research made use of data
  obtained from the High Energy Astrophysics Science Archive Research Center
  Online Service, provided by the NASA/Goddard Space Flight Center.  This
  work was supported by a grant of Russian Foundation of Basic Research 10-02-00492-a, NSh-5069.2010.2, and program of
  Presidium of RAS ``The origin and evolution of stars and galaxies'' (P-19).

\label{lastpage}
\end{document}